\documentstyle[twoside,fleqn,espcrc2]{article}

\newcommand{\AmS}{{\protect\the\textfont2
  A\kern-.1667em\lower.5ex\hbox{M}\kern-.125emS}}
\title{Back Reaction and Semiclassical Approximation of cosmological
       models coupled to matter}

\author{H. Wissowski\address{Institut f. Theor. Physik E, RWTH Aachen,
        52056 Aachen, Email: henning@physik.rwth-aachen.de}\thanks{H. W. thanks
the DFG Graduiertenkolleg for financial support.}
        and 
        H.A.\ Kastrup\address{Institut f. Theor. Physik E, RWTH Aachen, 52056
        Aachen,
        Email: kastrup@physik.rwth-aachen.de}}

\begin{document}

\begin{abstract}
Bianchi -I, -III, and FRW type models minimally coupled to a massive spatially
homogeneous scalar field (i.e. a particle) are studied in the framework of
semiclassical quantum gravity. In a first step we discuss the solutions 
of the corresponding equation for a Schr\"odinger particle propagating on a 
classical background.  \\
The back reaction of the Schr\"odinger particle on the classical metric is 
calculated by means of the Wigner function and by means of the expectation value
of the energy-momentum-tensor of the field as a source. Both methods in
 general lead to different results. 
\end{abstract}

\maketitle
\vspace{-4mm}
\noindent
\mbox{{\em To appear in: Proceedings of the Second Meeting on constrained 
Dynamics and Quantum Gravity}}
\noindent 
\mbox{{\em (Santa Margherita Ligure 1996)}; PITHA 97/14}

\section{Introduction}

Applying the well--known Dirac quantization procedure to general relativity 
coupled to a free massive scalar field leads to the corresponding 
Wheeler--DeWitt equation and to the diffeomorphism constraints. Expanding
the wave functional in powers of the gravitational constant yields a
semiclassical approximation  
\cite{kiefer1}\cite{singh1}, which describes the quantized matter field in 
a classical curved 
spacetime, i.e.
the quantized field propagate on a classical background defined
by the gravitational degrees of freedom. 
It is then natural to ask in which way the
background metric is influenced by the quantized field. Unfortunately 
no general procedure is yet known for calculating this kind of
back reaction from quantum
gravity.

\vspace{2.5mm}
The aim of this contribution is the following one:

\vspace{1mm}
a) A discussion of the semiclassical 
approximation for different types of minisuperspaces especially for
Bianchi -I, -III, and Friedman-Robertson-Walker type models minimally coupled
to a massive particle.  
\newpage

\vspace*{7mm}
b) To compare the results of two different approaches 
for calculating the back 
reaction in detail: The back reaction is calculated first with help of
Wigner's function and second by using the expectation value of the 
energy-momentum-tensor of the particle as a source.

\vspace{2.5mm}
The paper is organized as follows:

\vspace{1mm}
In chapter 2 we sketch the main features of semiclassical quantum gravity.

In chapter 3 the solutions of the corresponding Schr\"odinger 
equation of the quantized particle are given.

In chapter 4 we calculate and compare the back reaction on the metric by the
different methods 
mentioned above. 
 
\vspace{-1mm}
\section{Semiclassical Quantum Gravity}
\vspace{-1mm}

Starting from the Wheeler--DeWitt equation for minisuperspaces
\vspace{-1mm}
\begin{eqnarray}
\left[-h^{-1/2}G^2\hbar^2G_{abcd}\partial_{h_{ab}}\partial_{h_{cd}}
\right.&&\nonumber
\\
\left.
-\sqrt{h}(R
-2\Lambda)+GH_M\right]&&\!\!\!\!\!\!\!\!\!\!\!\!\Psi(h_{ab},\phi)=0,\label{1}
\end{eqnarray}
\vspace{-1mm}
where $G$ is $16\pi$ times the gravitational constant, $h_{ab}$ the homogeneous 
three-metric, $h$ its determinant, $R$ the three-dimensional Ricci scalar, 
$\Lambda$ the cosmological
constant, $G_{abcd}$ the metric on superspace,
$\partial_{h_{ab}}\!=\!\frac{\partial}{\partial h_{ab}}$ and 
$H_M$ the Matter-Hamiltonian 
for the massive spatially homogeneous field $\phi(t)$

\vspace{-1mm}
\begin{equation}
H_M=-\hbar^2 (2\sqrt{h})^{-1} \partial_{\phi}^2+
\sqrt{h}m^2\phi^2/2.
\end{equation} 
\vspace{-1mm}

We expand the wave function $\Psi$ in powers of $G$

\begin{equation}
\Psi=\exp\{\dot{\imath}[G^{-1}S_{-1}(h_{ab})+S_0(h_{ab},\phi)+\ldots]/\hbar\},
\end{equation}
where $S_{-1}$ is a solution of the Hamilton--Jacobi equation of pure
gravity. Defining the function $f(h_{ab},\phi)$ and the semiclassical 
time $t(h_{ab})$ by

\begin{equation}
f(h_{ab},\phi)=F(h_{ab})
\exp(\dot{\imath}S_0(h_{ab},\phi)/\hbar),
\end{equation}
\begin{equation}
\partial_{t}=G_{abcd}(\partial_{h_{ab}} S_{-1})
\partial_{h_{cd}},
\end{equation}
where $F(h_{ab})$ is determined by $S_{-1}$ \cite{kiefer1}\cite{singh1}, 
we obtain the Schr\"odinger equation on curved space 

\begin{equation}
\dot{\imath}\hbar\partial_{t} f(\phi,t)=H_M\left(\phi,t,
\partial_{\phi}\right)f(\phi,t).\label{Schr}
\end{equation}

Since we are dealing with a time dependent Schr\"odinger equation,
we use the Schr\"odinger inner-product for the wave function $f(\phi,t)$
\vspace{-1mm}
\begin{equation}
<f|f>=\int_{-\infty}^{+\infty}d\phi \,  f(\phi,t)f^*(\phi,t),\label{inner}
\end{equation}
\vspace{-1mm}
where $f^*$ denotes the complex conjugated of $f$.

\section{Explicit solvable Models}

Our starting point is the Schr\"odinger equation 
(\ref{Schr}) for  a massive
field. A redefinition of the semiclassical time $\partial_{\tau}\!
=\!2\sqrt{h}\partial_{t}$ yields
\begin{equation}
i\hbar\partial_{\tau}f=\left[\hbar^2\partial_{\phi}^2+\phi^2V(\tau)\right]f,
\quad V(\tau)=h(\tau)     
\label{fund}
\end{equation}
with a time-depent potential $V(\tau)$. Inserting the special ansatz
\begin{equation}
f_0=\exp\left(-\dot{\imath}\phi^2 \partial_{\tau}\ln[|y(\tau)|]/4\hbar-
\ln[|y(\tau)|]/2
\right)\!,\label{f0}
\end{equation}
yields for the unknown function $y(\tau)$ a second order ordinary differential
eq.
\begin{equation}
\frac{d^2 y(\tau)}{d\tau^2}=-V(\tau)y(\tau).\label{y}
\end{equation}
Assuming we have found a solution of (\ref{y}), we insert the unknown function
$\eta(\phi,\tau)$ defined by $f(\phi,\tau)\!=\!f_0(\phi,\tau)\eta(\phi,\tau)$ 
in eq.\ (\ref{fund}). 
After a Fourier-Transformation $\phi\rightarrow p, \eta\rightarrow 
\tilde{\eta}$ we conclude 
\begin{equation}
\partial_{\tau}\tilde{\eta}-p(\partial_{\tau}\ln(y))\partial_{p}\tilde{\eta}
-\dot{\imath}\hbar p^2 \tilde{\eta}=0.
\end{equation}
Thus solutions of (\ref{fund}) are given by
\vspace{-1mm} 
\begin{eqnarray}
f=&&\!\!\!\!\!\!\!\!\!\!f_0\int_{-\infty}^{+\infty}dp 
\, y\exp\{\dot{\imath}p\phi\}\nonumber
\\
&&\!\!\!\!\!\!\!\!\!\!\exp\{\dot{\imath}\hbar p^2 y^2
\int^{\tau}d\tilde{\tau}\,y^{-2}(\tilde{\tau})\}F(yp),
\end{eqnarray}
\vspace{-1mm}
with an arbitrary function $F(yp)$.
A complete set of eigenfunctions with respect to the inner-product 
(\ref{inner}) is given by
\vspace{-1mm} 
\begin{eqnarray}
f_n=&&\!\!\!\!\!\!\!\!\!\!\!y^{-1/2}H_n[\phi(4\dot{\imath}\hbar q(\tau))^{-1/2}
]\nonumber
\\
&&\!\!\!\!\!\!\!\!\!\!\!\exp[-\dot{\imath}\phi^2(\!\partial_{\tau}
\ln(y)\!)
/4\hbar 
\!+\!n\!\!\int^{\tau}\!\!\!\!d\hat{\tau}\,q^{-2}(\hat{\tau})/2],\label{Hermite}
\end{eqnarray}
\vspace{-1mm}
where $H_n$ are the Hermite polynomials of order n and $q(\tau)$ is defined by
\vspace{-1mm}
\begin{equation}
q(\tau)=y^2(\tau)\int^{\tau}d\tilde{\tau}\, y^{-2}(\tilde{\tau}).
\end{equation}
\vspace{-1mm}

Thus, the solutions of the Schr\"odinger eq.\ (\ref{fund}) can be calculated
from
the solutions $y(\tau)$ of eq.\ (\ref{y}). 
The solutions of eq.\ (\ref{y}) have to be chosen in such a way that 
$\Re(-\dot{\imath}\partial_{\tau}\ln(y))$ is positive i.e.\ the functions $f_n$\
(\ref{Hermite}) are normalizable.

\subsection{FRW Models}

A line element of the homogeneous and iso\-tropic Friedman--Robertson--Walker 
models is 
\begin{eqnarray}
ds^2=&&\!\!\!\!\!\!\!\!\!\!N^2(t)\,dt^2-a^2(t)[(1-kr)^{-1}\,dr^2\nonumber
\\
&&\!\!\!\!\!\!\!\!\!\!+\,r^2(\,d\theta^2+
\sin^2(\theta)\,d\phi^2)],
\end{eqnarray}
with $k\!=\!\pm1$ for the closed and for the hyperbolic universes respectively.
For vanishing $k$ we obtain the well--known DeSitter model, which 
describes a flat
three dimensional space.
For nonvanishing $k$ and cosmological constant $\Lambda$ we get a complicated
expression for the semiclassical time parameter $\tau$
\begin{eqnarray}
\tau=&&\!\!\!\!\!\!\!\!\!\!\{k^{-1}a^{-2}\sqrt{\Lambda a^2-k}
\nonumber\\
&&\!\!\!\!\!\!\!\!\!\!+\,\Lambda k^{-3/2} \arctan(\sqrt{\Lambda
k^{-1}a^2+1})\}/4.
\end{eqnarray}
Instead of this we introduce another parameter $x$ by $x\!=\!k\Lambda a^2$
and obtain from the eq.\ (\ref{y}) for $y(\tau)$ the well--known 
hypergeometric
differential equation
\begin{equation}
x(x-1)\partial_{x}^2y+(5x/2-2)\partial_{x}y+m^2y/(4\Lambda)=0.\label{hyper}
\end{equation}

For vanishing $\Lambda$, $k\!=\!-1$, we get with $\tau\!=\!a^{-2}/4$
Bessel's 
eq.\
\begin{equation}
9\Lambda\tau^3\partial_{\tau}^2y+m^2y=0.
\end{equation} 

In the DeSitter case with $k\!=\!0$, $\tau\!=\!a^{-3}/(6\sqrt{\Lambda})$ and 
$\Lambda>0$ eq.\ (\ref{y}) can be written as a special case of Bessel's
eq.\ namely Euler's eq.
\begin{equation}
9\Lambda\tau^2\partial_{\tau}^2y+m^2y=0.\label{euler}
\end{equation}
Thus for the DeSitter model the solutions of eqs.\ (\ref{euler}),(\ref{fund}) 
can be 
expressed by
elementary functions, depending on the value of $\Lambda/m^2$:
\begin{eqnarray}
\Lambda/m^2<4/9:\!\!\!\!\! &&y=\sqrt{\tau}\tau^{-\dot{\imath}k_0},
\\
\Lambda/m^2>4/9:\!\!\!\!\! &&y=\sqrt{\tau}
(k_1\tau^{k_0}+\dot{\imath}k_2\tau^{-k_0}),
\\
\Lambda/m^2=4/9:\!\!\!\!\! &&y=\sqrt{\tau}(k_1\ln(\tau)+\dot{\imath}k_2),
\end{eqnarray}
with $k_0\!=\!\sqrt{|m^2/(9\Lambda)\!-\!1/4|}$, $k_1,k_2\in R$ and $k_1k_2>0$
because of normalizability.

\subsection{Anisotropic Minisuperspaces}

Contrary to the FRW models discussed above each of the following two models 
possesses
two gravitational variables $z(t),b(t)$. 

\subsubsection{Bianchi--I with rotational symmetry}

The line element of the Bianchi--I model with rotational symmetry is 
\begin{equation}
ds^2\!=\!N^2(t)\,dt^2-z^2(t)\,dr^2-b^2(t)[d\theta^2+\theta^2\,d\phi^2].
\label{B-IL}
\end{equation}

Here we obtain for the semiclassical time parameter $\tau$ 
\begin{equation}
\tau=-\frac{2c_0}{b(4\Lambda b^2-3c_0^2z^2)}
\ln\left[\frac{4\Lambda b^2}{3c_0^2z^2}\right],\label{t}
\end{equation}
where $c_0$ is the separation constant of the solution  
\begin{equation}
S_{-1}=c_0z^2b+\Lambda b^3/(12c_0)
\end{equation}
of the Hamilton--Jacobi eq..
Inserting $\tau$ of (\ref{t}) in eq.\ (\ref{y}) we obtain 
\begin{equation}
\partial_{\tau}^2y+c_1\sinh^{-2}(c_2\tau)y=0,\label{B-I}
\end{equation}
with $c_1\!=\!3c_0^2m^2l^2/(32\Lambda)$ and $c_2\!=\!3c_0l/4$, where
$l\!=\!4\Lambda b^3/(3c_0^2)\!-\!z^2b$ has to be treated as a constant.

Eq.\ (\ref{B-I}) is related to a
hypergeometric one, which can be seen 
by a change of variables $w\!=\!1\!+\!\exp(2c_2\tau)$. This transforms
eq.\ (\ref{B-I}) into
\begin{equation}
w^2(1-w)\partial_{w}^2y-w^2\partial_w y+c_1y/c_2=0.
\end{equation}
Defining the function $\theta$ by $\theta(w)\!=\!y(w)w^{-c}$ with $c\!= \!1/2\!+
\!\sqrt{1/4\!
+\!c_1/c_2}$,
$\theta$ fulfills the hypergeometric differential eq..

\subsubsection{\hspace{-2mm} Bianchi-III \hspace{-1mm} with 
\hspace{-1mm} rotational 
\hspace{-1mm} symmetry}

Similar to the line element (\ref{B-IL}) above, we have
\begin{equation}
ds^2\!\!=\!\!N^2\!(t)dt^2\!\!-\tilde{z}^2\!(t)dr^2\!\!-\tilde{b}^2\!(t)
[d\theta^2\!\!
+\sinh^2\!(\theta)d\phi^2].
\end{equation}

Introducing the parameter 
\begin{equation}
x=(4c_3^2\tilde{z}^2+1)/(4c_3^2\tilde{z}^2-1),
\end{equation}
where $c_3$ denotes the separation constant of 
\begin{equation}
S_{-1}=c_3\tilde{z}^2\tilde{b}+\tilde{b}/(4c_3),
\end{equation} 
eq.\ (\ref{y}) yields
\begin{equation}
\partial_{x}(x^2-1)\partial_x y + c_4(1-x)^2y=0,\label{Wasser}
\end{equation}
with $c_4\!=\!2c_3^2m^2(4c_3^2\tilde{z}^2\!-\!1)\tilde{b}$ 
as a
constant.

This eq.\ (\ref{Wasser}) is known from the scattering problem for the hydrogen
molecule ion $H_2^+$ first solved by Jaff\'e \cite{Jaffe}.
\vspace{4mm}

The solutions of the Schr\"odinger equation (\ref{Hermite}) can be normalized
for each of the models above.

\section{Back Reaction}

Two different approaches for calculating the back reaction of the quantized matter field
on the classical metric are investigated.
First, using the expectation value of the energy--momentum tensor as a source
and second, with help of Wigner's function \cite{Wigner}.
We calculate and compare the results for different solutions of the 
Schr\"odinger eq.\ (\ref{fund}) for the DeSitter model.

\subsection{Using the energy-momentum tensor}

One of the most familiar definitions of back reaction is to replace 
$H_M$ by $<f|H_M|f>$ in the classical Hamiltonian constraint, where f
is the wave function (\ref{fund}).
This is equivalent to define
\begin{equation}
P=G^{-1}\partial_a S_{-1}+<f|\partial_a \beta(a,\phi)|f>, \label{rueck1}
\end{equation}
for the gravitational variable $a$ of the DeSitter model, where $P$ denotes the 
classical 
momentum conjugate
to $a$ and $\beta(a,\phi)$ the phase of the wave function $f$.

Inserting the unperturbed classical expression $P\!\!=\!\!-G^{-1}
\partial_t a^2$ ($N\!=\!1$) in eq.\ (\ref{rueck1})
we get a first order differential eq.\ for the perturbed
metric $\hat{a}(t)$. Since the 
perturbation is exact up to order $G$ only, we set 
$\hat{a}(t)\!=\!a(t)\!+\!Ga_1(t)$, where 
$a\!=\!\exp[\sqrt{\Lambda}(t\!+\!t_0)/2]$ is the unperturbed DeSitter metric. 

Given the wave function $f(\phi,\tau(a))$ we can calculate $a_1(t)$ explicitly.
For the ''groundstate'' $f_0\!= \!y^{-1/2} \exp(-\dot{\imath}\phi^2
\partial_{\tau}\ln(y)/(4\hbar))$ we obtain 
\begin{eqnarray}
\Lambda/m^2<4/9: \!\!\!\!\!\!\!\!\!\!&& a_1=-\hbar k_5^2a^{-2}[k_0+
1/(4k_0)],
\label{back1}
\\
\Lambda/m^2>4/9: \!\!\!\!\!\!\!\!\!\!&& a_1=-\hbar a^{-2}[a^{6k_0}k_6^2\!+
\!a^{-6k_0}k_7^2],
\label{back4}
\\
\Lambda/m^2=4/9: \!\!\!\!\!\!\!\!\!\!&& a_1=-\hbar a^{-2}[k_8^2(\ln (a))^2
\nonumber
\\
\!\!\!\!\!\!\!\!\!\!&&\quad\quad\,\, + 
\, k_9\ln(a) +
k_{10}],\label{back3}
\end{eqnarray}
with real constants $k_5,k_6,k_7,k_8,k_9, k_{10}$. 
In the massless case we have: $a_1\!=\! -\hbar k_{11}^2 a^{-5},$ 
$k_{11}\!\in \! R$. For large universes, $a\rightarrow\infty$, the perturbation 
is negligible $a_1\ll a,$ $\hat{a}\rightarrow a$.

If $f$ contains the Hermite polynomial $H_n$ we get 
\begin{equation}
a_1=-\hbar k_5^2 a^{-2}(2n+1)\{k_0+n!2^n/(4k_0)\}\label{back2}
\end{equation}
for $\Lambda/m^2\!<\!4/9$. Thus, the result (\ref{back1}) for 
the ''groundstate'' 
is only slightly modified. 
All the perturbations \ 
(\ref{back1}),(\ref{back4}),(\ref{back3}),(\ref{back2}) vanish in the 
''classical limit'' $\hbar\rightarrow0$.

To obtain the additional classical part of the back reaction of order
$G$, we
take the wave function
\vspace{-1mm} 
\begin{eqnarray}
f=\!\!\!\!\!\!\!\!\!\!&&y^{-1/2}\exp[-\dot{\imath}\phi^2\partial_{\tau}
\ln(y)/(4\hbar)
+\dot{\imath}
\tilde{c}\phi/(\hbar y) \nonumber
\\
\!\!\!\!\!\!\!\!\!\!&&+\,\dot{\imath}\tilde{c}^2\int^{\tau}d\grave{\tau}\,
y^{-2}(\grave{\tau})/\hbar],\label{backcl}
\end{eqnarray}
\vspace{-1mm}
and get
\vspace{-1mm}
\begin{equation}
a_1\!\sim\!-\tilde{c}_1^2a^{6k_0-2}\!-
\tilde{c}_2^2a^{-2}\!
-\tilde{c}_3^2a^{-6k_0-2}\!-\hbar \tilde{c}_4^2a^{6k_0-2}\!,
\nonumber
\end{equation}
\vspace{-1mm}
with real constants $\tilde{c}_1,\tilde{c}_2,\tilde{c}_3,\tilde{c}_4$.
Here the quantum corrections are given for large $a$.
Multiplying the wave function (\ref{backcl}) by Hermite polynomials
changes the coefficient $\tilde{c}_4^2$ 
of the quantum correction in $a_1$ only.

\subsection{Wigner's function}

Wigner's function $F_W$ 
is a generalization of a classical correlation
function on phase space \cite{Wigner}. 
$F_W(a,P,\phi,P_{\phi})$ depending on the gravitational, the matter variable
and their momenta is defined by
\vspace{-1mm}
\begin{eqnarray}
F_W\!=\!&&\!\!\!\!\!\!\!\!\!\!\!Gm^{-1}\!\!\int_{-\infty}^{\infty}\!\!\!du\,dv\,
f^{*}(a\!-\!G\hbar
u/2,\phi\!-\!\hbar v/(2m))
\nonumber
\\
&&\!\!\!\!\!\!\!\!\!\!\!e^{\!-\dot{\imath}(\!PuG+P_{\phi}v\!/m\!)}
\!f(\!a\!+\!\hbar
uG/2,\!\phi\!+\!\hbar v/\!(2m)\!).
\end{eqnarray}
Expanding $F_W$ in powers of $G$ leads in order $G^0$ to 
$F_W\!\sim\!\delta(P\!-\!G^{-1}\partial_a S_{-1})$ from which we obtain
the unperturbed classical expression $P\!=\!-G^{-1}\partial_{a}S_{-1}$ 
\cite{singh1}.
Integrating over $\phi,P_{\phi}$ we obtain in order $G$
\vspace{-1mm}
\begin{equation}
F_G\!=\!(\!\partial_a S_{-1}\!)^{\!-1}\!\!\!\int_{-\infty}^{\infty}\!\!\!\!\!d\phi
|f|^2\delta(\!P\!-\!G^{-1}\partial_a S_{-1}\!-\!\partial_a
\beta\!),
\end{equation}
\vspace{-1mm}
where $\beta(a,\phi)$ is the phase of $f$. The peaks
of $F_G$ yield the relation between $a$ and $P$ from which we 
obtain $a_1(t)$ by integration as in the case of the 
energy-momentum-tensor. 

For the ''groundstate'' we calculate
\begin{eqnarray}
\Lambda/m^2<4/9: \!\!\!\!\!\!\!\!\!\!&& a_1=-\hbar k_5^2k_0a^{-2},
\nonumber
\\
\Lambda/m^2>4/9: \!\!\!\!\!\!\!\!\!\!&& a_1\sim -\hbar a^{-2}(k_6^2a^{+6k_0}
\!-\!k_{11}^2a^{-6k_0}),
\nonumber
\\
\Lambda/m^2=4/9: \!\!\!\!\!\!\!\!\!\!&& a_1\sim -\hbar k_{12}^2a^{-2},
\quad \!k_{11},k_{12}\in R.
\end{eqnarray}
The last two expressions are given in leading order $a\!\gg\! 1$ only. The 
case
$\Lambda/m^2\!<\!4/9$ yields the same result as (\ref{back1}). 
$\Lambda/m^2\!>\!4/9$
agrees in leading order with (\ref{back4}). Only $\Lambda/m^2\!=\!4/9$
leads to a different back reaction compared with (\ref{back3}).
\\[2mm]

The quantum corrections induced by the wave function  
(\ref{backcl}) remain unchanged.

\end{document}